\documentclass{aa}
\usepackage{graphicx}
\newcommand{\be}{\begin{equation}}
\newcommand{\ee}{\end{equation}}
\newcommand{\lsi}{LSI~+61$^{\circ}$303}
\newcommand{\ls}{LS~5039}
\newcommand{\cg}{2CG~135+01}
\newcommand{\eg}{3EG~J0241$+$6103}
\begin{document}
\title{$\gamma$-ray emission from microquasars: a numerical model for \lsi}
\author{V. Bosch-Ramon, J. M. Paredes}
\institute{Departament d'Astronomia i Meteorologia, Universitat de Barcelona, Av. Diagonal
647, 08028 Barcelona, Spain\\\email{vbosch@am.ub.es; jmparedes@ub.edu}}
\authorrunning{Bosch-Ramon \& Paredes}
\titlerunning{$\gamma$-ray emission from \lsi}
\offprints{V. Bosch-Ramon \\ }
\abstract{We explore the possible association between the microquasar \lsi\ and the 
EGRET source \cg/\eg\ by studying, with a detailed numerical model, whether this system
can produce the emission and the variability detected by EGRET ($>$100~MeV) through 
inverse Compton (IC) scattering. Our numerical approach considers a population of relativistic
electrons entrained  in a cylindrical inhomogeneous jet, interacting with both the radiation
and the magnetic fields, taking into account the Thomson and Klein-Nishina regimes of
interaction. Our results reproduce the observed spectral 
characteristics and variability at $\gamma$-rays, thus strengthening the 
identification of \lsi\ as a high-energy $\gamma$-ray source. 
}  

\maketitle

\section{Introduction} \label{intro}

\lsi\ is a High Mass X-ray binary system whose optical counterpart is a bright (V$\sim$10.8) 
star of B0 V spectral type (Paredes \& Figueras \cite{Paredes&figueras86}). In fact, the 
companion star is a Be, and the compact object could be likely a neutron star 
(Hutchings \& Crampton \cite{Hutchings&crampton81}). Taylor et~al. 
(\cite{Taylor80}) found that \lsi\ presented strong radio outbursts each 
26.5 days, which was associated to the orbital period of the binary
system (Taylor \& Gregory \cite{Taylor&gregory82}). A four-year modulation
of the maximum flux during radio outbursts (Paredes \cite{Paredes87},
Gregory et~al. \cite{Gregory89}) was associated either to a
precession of a possible jet or to variations in the accretion rate (Gregory et~al.
\cite{Gregory89}). Optical observations carried out by Paredes \& Figueras 
(\cite{Paredes&figueras86}) detected variable emission at timescales similar to the radio ones 
and, from ROSAT data, Goldoni \& Mereghetti (\cite{Goldoni&mereghetti95}) found also X-ray 
variability on a timescale of days. The microquasar nature of \lsi\ was established when a 
relativistic jet structure was detected through VLBI observations of this source (Massi 
et~al. \cite{Massi01}). 

\lsi\ was proposed by Gregory \& Taylor (\cite{Gregory&taylor78}) as the  possible counterpart
of the high-energy $\gamma$-ray source \cg\  (Hermsen et~al. \cite{Hermsen77}, Swanenburg et~al.
\cite{Swanenburg81}),  which was also detected by
EGRET\footnote{http://cossc.gsfc.nasa.gov/egret} (\eg) (Kniffen et~al. \cite{Kniffen97}, Hartman
et al.  \cite{3rdEGRETC}). The proposed association between \lsi\ and the high-energy 
$\gamma$-ray source is still unclear due to the high uncertainty in position of EGRET  sources.
Nevertheless, no radio loud active galactic nucleus or strong radio pulsar is  known within the
error box of \eg, which includes \lsi\  (Kniffen et~al. \cite{Kniffen97}).   
Variability studies
can help us to reinforce the association between \lsi\ and \eg. Tavani et~al.
(\cite{Tavani98}) found that the $\gamma$-ray flux varied by a factor of 3. Also, 
Wallace et al. (\cite{Wallace00}) showed that in the EGRET viewing period 211.0 (1993 February
25--March 9) there was a $\gamma$-ray flare at an orbital phase around 0.5. Moreover, Massi
(\cite{Massi04b}) presented a variability analysis of the EGRET data, obtaining a $\gamma$-ray
period of 27.4$\pm$7.2 days, in agreement with the orbital one. The results of that work
also seem to suggest the presence of two peaks in the $\gamma$-ray flux: the first one would be 
in a phase around 0.2 (periastron passage, Casares et~al. \cite{Casares04}), which was not 
covered in the data studied by Wallace et~al. (\cite{Wallace00}), 
and the second one would be in a phase around 0.5 (like in the work of Wallace et~al. 
\cite{Wallace00}). 
It is worth mentioning that X-ray observations performed by ROSAT just six months 
before the EGRET viewing period 211.0 (1992 August--September) showed also an X-ray 
peak at an orbital phase around 0.5
(Peracaula \cite{Peracaula97}, Taylor et~al. \cite{Taylor96}). 
All these results strengthen the \eg/\lsi\ association. At very high-energy $\gamma$-rays,
Hall et~al. (\cite{Hall03}) gave upper limits for the emission associated to this source from 
observations performed with the Cherenkov telescope Whipple.

Several models have been proposed in order to explain the high energy emission of  this source
(e.g. Taylor et~al. \cite{Taylor96}, Punsly \cite{Punsly99}, Harrison et~al.  \cite{Harrison00},
Apparao \cite{Apparao01}, Leahy \cite{Leahy01}, Leahy \cite{Leahy04}). Two typical scenarios
have been proposed: a Be star  plus a non-accreting pulsar or an accreting compact object with
the same stellar  companion. The most of the models mentioned above focused on the IC scattering
of stellar  photons by relativistic electrons as the mechanism for generating X-rays and/or 
$\gamma$-rays. The work of Punsly (\cite{Punsly99}) deserves a particular  mention, since
Synchrotron Self Compton (SSC) scattering plays an important  role therein, like in the present
paper. We are interested in investigating whether \lsi\ is able  to generate the high-energy
$\gamma$-ray emission detected by EGRET. However, in the present  work, we have performed
accurate numerical calculations, taking into account the electron energy losses,
to obtain the spectral emission from external
Compton (EC) and SSC interactions in the Thomson and Klein-Nishina regimes.  Moreover, unlike
the pulsar wind shock model (first proposed by Maraschi \& Treves \cite{Maraschi&treves81} and
adopted also by, e.g., Harrison et~al. \cite{Harrison00} and  Leahy \cite{Leahy04}), we suppose
that the electrons are entrained within a jet, which  is ejected from the compact object. As a
matter of fact,  the discovering of relativistic radio jets, very similar to those observed in
other  microquasars (Mirabel \& Rodr{\'{\i}}guez \cite{Mirabel&rodriguez99}), gives strong
support to the accreting compact object scenario. 
It should be noted that leptonic jet models are
not the only way to explain emission at EGRET energies. For instance, in the context of 
a high-mass microquasar like \lsi,  
a hadronic jet model emitting $\gamma$-rays through pion-decay (see, i.e., Romero et~al.
(\cite{Romero03}) could also be applied. In such a case, neutrinos would be expected also, as
well as an EGRET spectrum harder than the one produced in the leptonic case, due to the much lower
rate of losses for protons than for electrons in the jet.

\ls\ (3EG~J1824$+$1514) (Paredes et~al. \cite{Paredes00}, 
%ull, olla amb les referencies
Rib\'o \cite{Ribo02}) is another
microquasar and  EGRET source candidate, sharing similar characteristics with \lsi: both sources 
are high mass X-ray binaries, both sources seem to harbor a neutron star (for \ls,  McSwain et~al.
\cite{Mcswain04}), both present persistent radio jets (for \ls, Paredes et~al. \cite{Paredes02};
for \lsi, see Massi et al. \cite{Massi01} and \cite{Massi04}), the apparent absence of inner disk
features in the  X-ray spectrum, their moderate levels of radio and X-ray emission (for \lsi, in
the quiescent state), their possible nature as high-energy $\gamma$-ray  emitters, and that both
of them are likely fed by wind from the primary (for \lsi, unless perhaps around periastron). Our
model, previously applied to \ls\ (Bosch-Ramon \& Paredes \cite{Bosch-Ramon&paredes04}), has in
\lsi\ a new interesting object of  application.

This paper is organized as follows: in Sect.~\ref{thephy} we explain 
briefly our model, the application of the model to \lsi\ and its results are presented 
in Sect.~\ref{app}, and the discussion of these results is developed in Sect.~\ref{disc}.

\section{The model} \label{thephy}

In this model, we assume that the leptons in the jet dominate the radiative processes related to
the $\gamma$-ray production. The relativistic population of electrons, already accelerated 
and flowing away into the
jet, is exposed to external photons as well as to the synchrotron photons emitted by 
the electrons, 
since we take into account the magnetic field in our model.
The $\gamma$-ray emitting region, the $\gamma$-jet, is assumed to be closer to the compact
object than the observed radio jets. This $\gamma$-jet is supposed to be short
enough to be considered cylindrical. The magnetic field ($B_{\gamma}$) has been taken 
to be constant, as an average along the jet.
We have included the interaction
between the relativistic electrons and both the magnetic and the radiation fields. The energy
losses of the relativistic leptonic plasma within the $\gamma$-jet are mainly due to synchrotron emission, SSC scattering, and EC scattering. Due to the
importance of the losses, the electron energy distribution density along the $\gamma$-jet model
varies significantly, and in this sense the $\gamma$-jet is non-homogeneous.
The $\gamma$-jet is studied by splitting it into cylindrical transverse cuts or slices. The
size of the slices has to be suitable in order to get almost homogeneous physical conditions
within each one (energy densities of the radiation and the electrons). Regarding IC interaction,
we have used the cross-section of Blumenthal \& Gould (\cite{Blumenthal&gould70}):
$\sigma(x,\epsilon_0,\gamma_{\rm e})$, which takes into account the low- and the high-energy
regimes of interaction (i.e. the Thomson and Klein-Nishina regimes), $\epsilon_0$ is the seed 
photon energy, $\gamma_{\rm e}$ is the scattering electron Lorentz factor, and $x$ is actually 
a function
which depends on both of the former quantities and on the scattered photon energy ($\epsilon$).

The electron distribution is assumed to be initially a power law
($N(\gamma_{\rm e})\propto \gamma_{\rm e}^{-p}$, where $\gamma_{\rm e}$ is the electron Lorentz 
factor), which evolves under the
conditions imposed by the magnetic and the radiation fields. Thus, the electron distribution
function of a certain slice ($N(\gamma_{\rm e},z)$) depends on both the distance to the compact
object ($z$) and $\gamma_{\rm e}$. The components of the total seed photon radiation field 
($U(\epsilon_0,z)$) are any present external radiation field ($U_{\rm ext}(\epsilon_0,z)$)  
and the synchrotron radiation field produced by the relativistic electrons within the jet, all 
of them in the reference frame of the jet (for the external photon fields, see Dermer 
\& Schlickeiser \cite{Dermer&schlickeiser02}).
 
The free parameters of the model are $B_{\gamma}$ and the
maximum electron Lorentz factor at the slice closest to the compact object ($\gamma_{\rm
e0}^{\rm max}$). The leptonic kinetic luminosity or leptonic jet power ($L_{\rm ke}$) is set
free also, and it is scaled with the observed luminosity, in order to reproduce the observations.

The luminosity per energy unit in the jet's reference frame ($L_{\epsilon}$) is presented in 
Eq.~\ref{eq:lumicSRj}. The photon flux per energy unit or spectral photon 
distribution\footnote{In our previous paper (Bosch-Ramon \& Paredes
\cite{Bosch-Ramon&paredes04}), the factor $\epsilon_0$ in the denominator of the integrand in
Eq.~\ref{eq:lumicSRj} was inadvertently left out (see Eq.~12 in that work), although correctly
included in the calculations. Also, the notation related to calculation reference frame has been 
clarified.} 
in the reference frame of the observer($I'_{\epsilon'}$) is shown in 
Eq.~\ref{eq:lumicSRobs}. The magnitudes with ($'$) are in the observer reference frame.

\begin{eqnarray}
L_{\epsilon}&=&\epsilon\sum^{z_{\rm max}}_{z_{\rm min}} V_{\rm slice}(z)
\int^{\epsilon_0^{\rm max}(z)}_{\epsilon_0^{\rm min}(z)}
\int^{\gamma_{\rm e}^{\rm max}(z)}_{\gamma_{\rm e}^{\rm min}(z)}
\frac{U(\epsilon_0,z)}{\epsilon_0}\nonumber \\&&
\times N(\gamma_{\rm e},z)
\frac{d\sigma(x,\epsilon_0,\gamma_{\rm e})}{d\epsilon}
d\gamma
d\epsilon_0
\label{eq:lumicSRj}
\end{eqnarray}

\be
I'_{\epsilon'}=\frac{\delta^{2+p}}{4\pi D^2 \epsilon'}L_{\epsilon'}
\label{eq:lumicSRobs}
\ee
where $\delta$ is the Doppler factor of the jet, $D$ is the distance from \lsi\ to
the observer, and $V_{\rm slice}(z)$ is the volume of the slice at a distance $z$ 
from the compact object. For further details of the model, see  Bosch-Ramon \& Paredes
(\cite{Bosch-Ramon&paredes04}).

\section{Application of the model to \lsi} \label{app}

In this section, we explore the possible nature of \lsi\ as a high-energy $\gamma$-ray emitter.
We apply our model going through different magnetic field strengths and maximum electron Lorentz
factors in order to estimate the best parameter values to reproduce the EGRET data. Afterwards,
we study the implications of the orbital eccentricity on the stellar photon density and the 
accretion rate, and how it yields variability at high-energy $\gamma$-rays. 
We note that in this work we have assumed that the external photon field is solely due to 
the stellar radiation field ($U_{\rm star}(\epsilon_0,z$)), taken to be a black-body. 
We have not considered other possible sources of external photons like a disk or a corona. For a 
semi-analytical model of an inhomogeneous jet accounting for all the external photon fields, 
see Bosch-Ramon et al. (\cite{Bosch-Ramon04}).

\subsection{Parameter election for \lsi} \label{par}

\begin{table*}[t!]
\begin{center}
\caption[]{Parameter values.}
\begin{tabular}{l c c c c c}
\noalign{\smallskip} \hline \hline \noalign{\smallskip} Parameter &
Description & derived values from observations \cr\noalign{\smallskip} \hline
\noalign{\smallskip}
$\Gamma_{\rm jet}$ & jet Lorentz factor & 1.25 \cr
$\theta$ & angle between the jet and the observer line of sight & $30^{\circ}$ \cr
$v_{\rm jet}$ & jet velocity & 0.6$c$ \cr
$a$ & orbital semi-major axis & $5\times10^{12}$~cm \cr
$e$ & orbital eccentricity & 0.7 \cr
$D$ & distance to the observer & 2 kpc \cr
$L_{\rm star}$ & star total luminosity & $2\times10^{38}$~erg~s$^{-1}$ \cr
$I_{\rm >100~MeV}$ & photon flux at the EGRET band &
$8\times10^{-7}$~photon~cm$^{-2}$~s$^{-1}$ \cr
$\Gamma$ & photon index at the EGRET band & 2.2
\cr \noalign{\smallskip} \hline \noalign{\smallskip}
Parameter & Description & adopted values
\cr \noalign{\smallskip} \hline \noalign{\smallskip}
$R_{\gamma}$ & $\gamma$-jet radius & $10^7$~cm \cr
$L_{\rm ac}$ & accretion disk luminosity & $10^{-9}~M_{\odot}~c^2$~yr$^{-1}$
\cr\noalign{\smallskip}
\hline\end{tabular}
\label{param}
\end{center}
\end{table*}

\lsi\ is an X-ray binary system located at an estimated distance of 2~kpc (Frail \& Hjellming
\cite{Frail&hjellming91}). The  bolometric luminosity of the companion star has been taken to be
$L_{\rm star}\sim2\times10^{38}$~erg~s$^{-1}$.  The more accurate value for the radio period,
$P=26.4960 \pm 0.0028$~days, assumed to be equal to the orbital period, was determined by
Gregory (\cite{Gregory02}), and the eccentricity is taken in concordance with Mart{\'{\i}} \&
Paredes (\cite{Marti&paredes95}) and Casares et~al. (\cite{Casares04}): $e\sim 0.7$. The orbital
semi-major axis  will be adopted as it is usual in the literature: $a=5\times10^{12}$~cm.  We
define the distance between the companion star and the compact object as $R_{\rm orb}$.  From
MERLIN observations carried out by Massi et~al. (\cite{Massi04}),  the jet velocity ($v_{\rm
jet}$) is taken to be 0.6$c$. As it is stated in that paper, the jet could be precessing,
implying a possibly strong variation of $\theta$. A typical value for $\theta$ of $\pi/6$ has
been adopted. This $v_{\rm jet}$ implies a jet Lorentz factor of $\Gamma_{\rm jet}=1.25$. Also,
according to radio observations (Ray et~al. \cite{Ray97}), the radio spectral index ($\alpha$)
can vary from $-$0.4  to 0 (where the flux density is $F_{\nu}\propto \nu^{\alpha}$),  with
$\alpha\sim -0.4$ in the quiescent state. These values of $\alpha$, following the simple
relationship  $p=1-2\alpha$ (only valid in the synchrotron optically thin regime), would imply a
value of $p$ in the range from 1 to 1.8, being the second number its value in the quiescent state. We will
try values of $p$ around its value in the quiescent state in our study. The observed spectrum
above 100~MeV has been obtained from the third EGRET catalogue (Hartman et~al.
\cite{3rdEGRETC}). The total observed photon flux at energies higher than 100~MeV is about
$8\times10^{-7}$~photons~cm$^{-2}$~s$^{-1}$, and the observed photon index is $\Gamma=2.2\pm0.1$
(photon flux per energy unit: $I_{\epsilon}\propto \epsilon^{-\Gamma}$). All these parameters
have been summarized in Table~\ref{param}. 

To fix the $\gamma$-jet radius ($R_{\gamma}$), we have imposed that it should be
at least of about few electron Larmor radii ($\sim 10^6$~cm), in order to keep those electrons 
confined inside the jet. Thus, $R_{\gamma}$ has been taken to be $10^7$~cm. To determine
the power of the leptonic jet, we need to fix also the accretion luminosity of the disk (Falcke \&
Biermann \cite{Falcke&Biermann96}). A typical value for a microquasar with a high mass stellar
companion can be $L_{\rm ac}\simeq 10^{-9}~M_{\odot}~c^2$~yr$^{-1}$. Both prior parameters can
also be found in Table~\ref{param}. $L_{\rm ke}$ will be fixed through comparison between the
observed fluxes and the model. From the EGRET energy range, and the involved seed photon and
electron energies in the scattering, the initial maximum Lorentz factor of the electrons should
be about $10^4$. Also, similar values for the maximum Lorentz factor were estimated from millimiter
observations by Paredes et~al. (\cite{Paredes00b}). A more accurate value for 
$\gamma_{\rm e0}^{\rm max}$ will be found
when trying to reproduce the observed spectrum slope. Regarding $B_{\gamma}$, we will study our
model behavior along a wide range of magnetic field values, from values of 1~G (close to the magnetic
field strengths at the radio emitting zone, see Massi et~al. \cite{Massi93}) to 100~G.

\subsection{Results} \label{res}

\begin{figure}
\resizebox{\hsize}{!}{\includegraphics{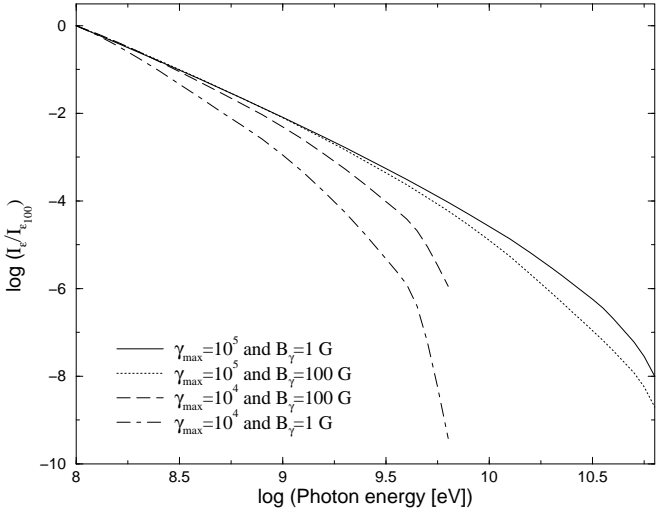}} \caption{Computed spectral photon 
distribution normalized to the photon flux value at 100~MeV for different values of the 
magnetic field and the maximum electron Lorentz factor. The solid and dotted lines represent 
$I'_{\epsilon'}$ for two different $B_{\gamma}$: 1~and~100~G, respectively, and a 
$\gamma_{\rm e0}^{\rm max}$ of $10^5$. The dot-dashed and dashed lines represent 
$I'_{\epsilon'}$ for two different $B_{\gamma}$: 1~and~100~G, respectively, and a 
$\gamma_{\rm e0}^{\rm max}$ of $10^4$.}
\label{figcases}
\end{figure}

We have computed the normalized spectral photon distributions for two different values 
of $B_{\gamma}$ and $\gamma_{\rm e0}^{\rm max}$ (see Fig.~\ref{figcases}). For 
$\gamma_{\rm e0}^{\rm max}=10^4$ and $B_{\gamma}$=100~G, the computed spectral 
photon index above 100~MeV is 2.4, which is steeper than the
observed one (2.2$\pm 0.1$). In case that the magnetic field strength is lower, 
the spectrum will get softer. For $\gamma_{\rm e0}^{\rm max}=10^5$ and 
$B_{\gamma}$=100~G, the spectrum becomes more similar to the one observed in the 
EGRET energy range. Now, in case that the magnetic field strengh is lower the 
spectrum will get harder. About the electron power law index, we have found 
that a good value might be around 1.7, and higher values of $p$ would imply 
a calculated photon index different from the one obtained from EGRET data 
(about 2~$\sigma$ or more for $p$ higher than 2). We want to remark that $p$ 
is the power-law of the initial electron energy distribution, and the spectral 
softening is a natural consequence of the introduction of losses in our model; it is 
not imposed a priori. 
Due to the lack of data beyond 10~GeV, we cannot still give a proper upper 
limit for $\gamma_{\rm e0}^{\rm max}$.

\begin{figure}
\resizebox{\hsize}{!}{\includegraphics{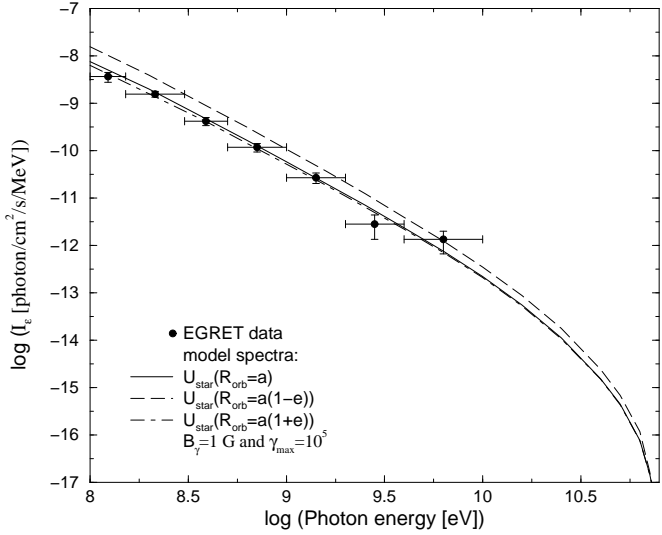}} \caption{Computed spectral photon 
distribution above  100~MeV plotted  with the EGRET data points. Only changes in the stellar
photon density have been taken into account and, due to the {\it low} magnetic field,
EC dominates. $p$ is taken to be 1.7, $\gamma_{\rm e0}^{\rm max}=10^5$,
and $B_{\gamma}$=1~G. There are plotted the
computed $I'_{\epsilon'}$ for different orbital radii: $a$ (solid line), the distance
at the periastron passage ($a(1-e)$, dashed line), and the distance at the apastron
passage ($a(1+e)$, dotted line).}
\label{boncas1}
\end{figure}

\begin{figure}
\resizebox{\hsize}{!}{\includegraphics{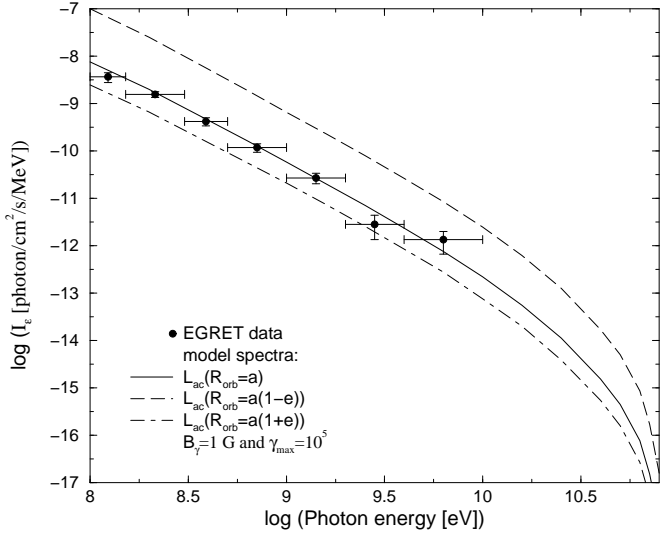}} \caption{Same as in
Fig.~\ref{boncas1} but now changes in both the stellar photon density and the accretion
rate have been taken into account.}
\label{boncas2}
\end{figure}

\begin{figure}
\resizebox{\hsize}{!}{\includegraphics{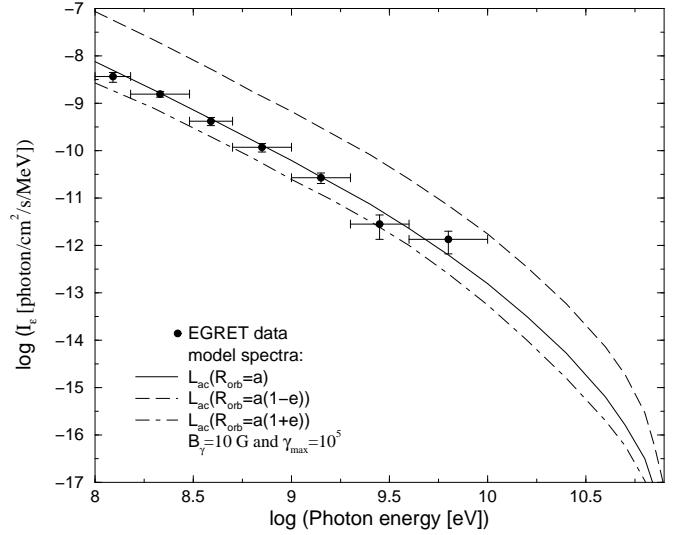}} \caption{Same as in
Fig.~\ref{boncas2} but with $B_{\gamma}$=10~G.}
\label{boncas3}
\end{figure}

The values of the model parameters that reproduce properly the observations are shown in Table~\ref{parval}. We have taken separately the case in which the dominant source of seed
photons is the companion star (i.e. $B_{\gamma}$=1~G, see Figs.~\ref{boncas1}~and~\ref{boncas2})
and the case in which the dominant source of seed photons is the synchrotron process
within the jet (i.e. $B_{\gamma}$=10~G, see Fig.~\ref{boncas3}). Also, in order to study the implications on variability of the orbital eccentricity at the
EGRET energy range, we have calculated the spectral photon distribution at an orbital 
distance equal to the orbital semi-major axis, at the periastron
passage and at the apastron passage. However, in a first and simpler situation
(Fig.~\ref{boncas1}), only the variations in the stellar photon density have been taken into
account. In the other two cases (Figs.~\ref{boncas2}~and~\ref{boncas3}), accretion variation has
been added. Following the accretion model of Bondi \& Hoyle (\cite{Bondi&hoyle44}),
the accretion rate has been taken to be proportional to the density of the medium surrounding the compact
object and further dependences have been neglected here. We have assumed also
that the ambient density decreases like $1/R_{\rm orb}^2$. 

It is worth noting that, although for the two magnetic field strengths quoted before the spectral
photon distribution is quite similar, the physical origin of the seed photons and the length of
the $\gamma$-jet are different.  When $B_{\gamma}$=1~G, the length reached by the jet's electrons
emitting at 100~MeV by IC process (the $\gamma$-jet length) is of about one astronomical unit 
whereas, for $B_{\gamma}$=10~G, the length of such a jet is roughly one hundred times smaller.

\begin{table}
\begin{center}
\caption[]{Two sets of free parameters used to reproduce the observations.}
\begin{tabular}{l c c c c
c}\noalign{\smallskip} \hline \hline \noalign{\smallskip}
Parameter & dominant EC & dominant SSC
\cr\noalign{\smallskip} \hline \noalign{\smallskip}
$B_{\gamma}$ & 1~G & 10~G \cr
$L_{\rm ke}$ & $10^{35}$~erg/s & $3\times10^{35}$~erg/s \cr
$\gamma_{\rm e0}^{\rm max}$ & $10^5$ & $10^5$
\cr\noalign{\smallskip}
\hline\end{tabular}
\label{parval}
\end{center}
\end{table}

\section{Discussion} \label{disc}

Our model is able to reproduce the observed data, supposing that the emission detected by EGRET
comes from a compact cylindrical jet close to the compact object. In order to reach the 
observed levels of emission, a $L_{\rm ke}$ of $10^{35}$~erg/s for $B_{\gamma}$=1~G, or of
$3\times10^{35}$~erg/s for $B_{\gamma}$=10~G, is necessary. These  values for the leptonic jet
power are well within the limits given in the work of Falcke \& Biermann
(\cite{Falcke&Biermann96}), using our adopted $L_{\rm ac}$. Regarding $\gamma_{\rm e0}^{\rm
max}$, for a value of  $10^5$,  the model reproduces properly the observations. Otherwise,
future instruments covering EGRET and higher energy ranges (i.e.
AGILE\footnote{http://agile.mi.iasf.cnr.it}, GLAST\footnote{http://glast.gsfc.nasa.gov},
HESS\footnote{http://www.mpi-hd.mpg.de/hfm/HESS/HESS.html},
MAGIC\footnote{http://hegra1.mppmu.mpg.de/MAGICWeb/}) will allow to get a better constraint  on
$\gamma_{\rm e0}^{\rm max}$, as well as will permit to check whether the predicted  spectral
shape follows the real source spectrum. Also, assuming that the electron energy distribution
function is represented roughly by a power-law,  we have found that $p$ should be of about 1.7,
which is also  in agreement with radio observations in the quiescent state. This fact seems to
suggest  that both radiation types (radio and $\gamma$-rays) come from the jet. Our model also
predicts   that, during a flare, variations in both the intensity and the slope of the
spectrum  at higher energies should be detected in correlation with the ones observed at  radio
frequencies. This is not necessarily in contradiction with the observed shift in time  of the
emission peaks at different energy bands, which have been linked to  changes in the environment
and within the jet itself (Mart{\'{\i}} \& Paredes \cite{Marti&paredes95}, Strickman et~al.
\cite{Strickman98}, Gregory et~al. \cite{Gregory99}) and whose study is beyond the scope 
of this work. Otherwise, good timing and
spectral resolution observations at high-energy $\gamma$-rays  and coordinated observations at
different energy bands are needed in order to find  out whether both the low- and the
high-energy emission have the same origin. We will not go, for the moment, into the study of
the emission at frequencies below  the high-energy $\gamma$-rays, being necessary to extend the
modelisation of the jet up to  bigger scales. A further comment can be done, regarding the role
of the magnetic field in our model: unlike in some other developed models before, $B_{\gamma}$
could play an important role  in the generation of seed photons in the high energy IC emission.
It is worth also mentioning  that the upper limits of emission at hundreds of GeV (Hall et~al.
\cite{Hall03}) would be in  agreement with the softening of the computed spectral photon
distribution above $\sim$10~GeV.  This softening above $\sim$10~GeV would make the source
extremely faint at hundreds of GeV and  above. 
The previous point and a spectrum at
X-rays harder than at EGRET energies, likely due to the electron energy losses, seem to point to a
dominant leptonic radiative mechanism in the jet (IC) instead to a hadronic one.

Regarding variability, our model predicts that important variations of the flux might occur
along the eccentric orbit. For the sake of simplicity, we have contemplated two cases. The first
one  accounts for the variation of the EC photon flux along the orbit due only to the changes in
the stellar photon density because of the eccentricity (see Fig.~\ref{boncas1}). This is only 
relevant if the SSC effect is not significant. The second one takes into account  the variation
of the IC photon fluxes due to the changes in both the stellar photon density and the accretion
rate because of the orbital eccentricity (see Figs.~\ref{boncas2}~and~\ref{boncas3}). The
high-energy $\gamma$-ray emission variability is clearly dominated by the accretion rate
variations along the orbit. As it is mentioned in Sect.~\ref{intro}, the high-energy
$\gamma$-ray emission of \eg\ varies a factor of 3. Also, all of the computed spectral photon
distributions present fluxes that vary within a range of 2--30 times.  Due to EGRET timing
sensitivity, the periodic outburst at $\gamma$-rays  associated to the periastron passage could
not be sampled properly though the detected {\it averaged} emission could explain the observed
variability of \eg, being also related to the strong radio outbursts. Such a $\gamma$-ray
outburst would  be detected smoother and earlier in orbital phase than the radio one. If this is
true, more  sensitive timing observations at high-energy $\gamma$-rays along the orbit will find
larger variability than the previously found. 
It is interesting to note that our model predicts a peak in the periastron at 0.2 
(as it is found in the work of Massi \cite{Massi04b}), when accretion rate is expected to 
be higher. The second peak at phase 0.5, mentioned in the
Sect.~\ref{intro}, can be explained in the context of an eccentric orbit where the interaction 
between the compact object and the wind of the stellar companion produces an 
increase in the accretion rate under certain conditions (Mart{\'{\i}} \& Paredes 
\cite{Marti&paredes95}). 
This would produce also a variability effect on the $\gamma$-ray flux like the one 
showed in Figs.~\ref{boncas2}~and~\ref{boncas3}.

Short timescale variability could be produced by precession (see, for instance, Kaufman Bernad\'o et~al.
\cite{Kaufman02}). In Massi et~al. (\cite{Massi04}), the authors  found evidence of precession
of the \lsi's jets on timescales of a few days, similar to the shorter variability timescales
found at high-energy $\gamma$-rays by Tavani et~al. (\cite{Tavani98}) for \cg. We have computed
the fluxes for different values of the angle $\theta$, from 0 to $\pi$/2,  obtaining variations
in the intensity of the IC emission of almost two orders of magnitud.  This gives us just an
upper limit for the intensity changes of the $\gamma$-ray emission due to precession, in
agreement  with observed variability at these energy ranges. 

Regarding the effect of the magnetic field on the length-scale of the $\gamma$-jet, for
$B_{\gamma}$ above 10~G and due to the strong energy losses induced by the SSC effect,
the electrons might need to be reaccelerated significantly after leaving the $\gamma$-jet 
to reach the observed radio jet. However, for $B_{\gamma}$ lower than 10~G, 
reacceleration might not be necessary in regions closer to the compact object 
than the radio jet. It is remarkable that higher magnetic fields imply more 
seed photons though higher energy losses as well, and this is the reason why the 
leptonic jet power requirements are slightly different depending on the dominant 
mechanism of $\gamma$-ray emission, EC or SSC scattering (see Table~\ref{parval}).

Finally, comparing the results obtained applying our model to \lsi\ with the ones obtained  in
Bosch-Ramon \& Paredes (\cite{Bosch-Ramon&paredes04}) for LS~5039, we want to note the strong
similarities shown by both sources. Nevertheless,  it is remarkable that the jet power in  the
first case is ten times smaller than in the second one. This is related to the fact that both
sources present different Lorentz factors. The Lorentz factor of the jet in \lsi\ is mildly
relativistic but higher than the Lorentz factor of the jet in LS~5039 (1.25 and 1.02, 
respectively). This higher Lorentz factor has two effects. The first one is the increase of the
observed flux due to the Doppler boosting, needing less kinetic power in the jet to explain the 
observed levels of emission. The second effect is an increase of the stellar seed photon density
in the reference frame of the jet of \lsi. This makes the photon field density of the stellar 
companion of \lsi (a B0 V star) in the reference frame of the jet to be similar to the one 
in the case of LS~5039 (a ON6.5~V((f)) star). Therefore, in order to know which source of seed 
photons can be dominant, is important also to have a good knowledge of both the jet kinematic 
characteristics and the angle between the jet and line of sight.

\begin{acknowledgements}
We are grateful to Marc Rib\'o and Gustavo E. Romero for their useful
comments and suggestions. V.B-R. and J.M.P. acknowledge partial support by DGI of the Ministerio
de Ciencia y Tecnolog{\'{\i}}a (Spain) under grant AYA-2001-3092, as well as additional support
from the European Regional Development Fund (ERDF/FEDER). During this work, V.B-R has been
supported by the DGI of the Ministerio de Ciencia y Tecnolog{\'{\i}}a (Spain) under the
fellowship FP-2001-2699.
\end{acknowledgements}

{}

\end{document}